\documentclass[pdflatex,sn-mathphys-ay,iicol]{sn-jnl}

\usepackage{pstool}
\usepackage{graphicx} 
\usepackage{subcaption}
\usepackage{amsmath,amssymb,amsfonts,amsbsy,bbm,bm}
\usepackage{stmaryrd}
\usepackage{array} 
\usepackage{multirow} 
\usepackage{svg}

\newtheorem{remark}{Remark}%

\newcommand{\tensor}[1]{\mbox{\boldmath ${#1}$}}
\newcommand{\IntB}{\int_{\mathcal{B}}\ }
\newcommand{\IntSB}{\int_{\partial\mathcal{B}}\ }

\theoremstyle{thmstyleone}%
\theoremstyle{thmstyletwo}%

\theoremstyle{thmstylethree}%

\begin{document}

\title[Article Title]{Deblurring structural edges in variable thickness topology optimization via density-gradient-informed projection}


\author*[1]{\fnm{Gabriel} \sur{Stankiewicz}}\email{gabriel.stankiewicz@fau.de}
\author[1]{\fnm{Chaitanya} \sur{Dev}}\email{chaitanya.dev@fau.de}
\author[1]{\fnm{Paul} \sur{Steinmann}}\email{paul.steinmann@fau.de}

\affil*[1]{\orgdiv{Institute of Applied Mechanics}, \orgname{Friedrich-Alexander-Universit\"at Erlangen-N\"urnberg}, \orgaddress{\street{Egerlandstr. 5}, \city{Erlangen}, \postcode{91058}, \state{Bavaria}, \country{Germany}}}


\abstract{Variable thickness topology optimization (VTTO) is a potent methodology for designing high-performance, high-stiffness sheet structures. However, this method frequently encounters two primary challenges: 1) the formation of undesirable low-thickness regions, which present manufacturing difficulties, and 2) the blurring of structural edges. This blurring is an artifact inherent to the regularization filters required for well-posedness. This paper proposes solutions to address both challenges. First, to mitigate low-thickness regions, we introduce a robust, combined approach. This strategy utilizes a SIMP-based penalization and an updated projection method, which effectively suppresses nearly all low-thickness domains. Second, the main contribution of this work is a novel method to deblur structural edges, termed the density-gradient-informed (DGI) projection. This projection utilizes local density gradient information. It selectively applies a strong projection in high-gradient regions (i.e., structural edges) to restore sharpness, while minimally affecting low-gradient regions within the structure's interior. Numerical examples demonstrate that the DGI projection successfully deblurs the structural edges, restoring a distinct solid-void transition, while preserving the internal form. Most importantly, this significant improvement in edge definition is achieved with a negligible impact on the final structural compliance. This establishes the DGI projection as a non-invasive and effective regularization tool for enhancing VTTO designs.}




\maketitle

\section{Introduction}\label{sec:intro}

Since Michell's classical theory (\cite{michell1904lviii}) of optimal frame structures, structural optimization and, in particular, topology optimization (TO) has developed into a powerful tool for creating structures with high performance and manufacturing feasibility. Early frameworks such as homogenization-based TO (\cite{bendsoe1988generating}), density-based methods (\cite{bendsoe1989optimal}), and level-set techniques (\cite{wang2003level}) shaped the field, with the density-based SIMP (Solid Isotropic Material with Penalization) approach becoming widely chosen because of its simple formulation and integration into commercial FEM tools. By penalizing intermediate densities ($p > 1$), SIMP favors black-and-white designs and prevents non-physical grey regions. On the contrary, as early as in the 1970s, \cite{rossow1973finite} proposed a two-dimensional optimization framework with variable thickness sheet (VTS) that bears a close relation to density-based methods. 

When no penalization is applied in the SIMP approach ($p = 1$), the interpretation of the design variables changes: intermediate values can be understood as thicknesses, as in the VTS optimization of \cite{rossow1973finite}. Variable thickness topology optimization (VTTO), a specific case of the VTS method representing a density-based or homogenization-based approach with possible topological changes, received little interest in its early phase, while immense developments in (penalized) TO methods were achieved. Characteristic of the penalized TO methods is the formation of visually appealing truss-like features. The obtained structures often mimic biological tissues, suggesting a valid comparison with nature's tendency to evolve and optimize its creations, further encouraging scientific developments in penalized TO methods. On the other hand, the VTTO approach renders visually less appealing sheet-like structures. However, the lack of penalization of intermediate densities enlarges the design space and can yield solutions of higher stiffness compared to penalized approaches. Only recently, \cite{sigmund2016non} demonstrated a comparative study between the (thin) truss-like and sheet-like structures, clearly indicating the superior stiffness of sheet-like structures.

The renewed interest in VTTO is also linked to the dehomogenization method, where coarse variable thickness fields are post-processed into high-resolution truss-like structures (\cite{groen2018homogenization, larsen2018optimal}), or multi-patch microstructures in a level-set setting (\cite{li2018topology}). Moreover, other VTS approaches have been extended to plates and shells (\cite{zhao2017stress, meng2022shape}), multimaterial designs (\cite{banh2019topology, nguyen2022multi}), and composites (\cite{stegmann2005discrete, sorensen2014dmto}). Advanced formulations couple thickness with shape or material optimization (\cite{meng2022shape, sjolund2018new, kashanian2021novel}).

The resurgence of VTTO highlighted both the potential and the challenges of such formulations. \cite{kandemir2018topology} compares a variety of approaches to obtain VTTO structures including the application of penalized TO, while \cite{yarlagadda2022solid} proposed the SIMTP extension with nodal thickness variables. However, one of the major challenges arising in VTTO is the formation of extremely thin sheets, which are sensitive to buckling, hard to interpret, and difficult to manufacture. For instance, \cite{groen2018homogenization} and \cite{larsen2018optimal} introduced a specialized smoothed Heaviside projection function that suppresses low-thickness regions for the dehomogenization step. \cite{giele2021approaches} introduced two strategies inspired by the cut element method, in which an additional level-set field is introduced that truncates low-thickness regions, \cite{geoffroy2022coupled, pozo2023minimum} suggested alternative penalization laws to avoid thin parts, while \cite{endress2023designing} applied a SIMP-like rule to suppress thicknesses below a threshold. All these contributions aimed at stable, production-ready designs without fragile sheet-like parts. In this work, we propose a generalized approach to avoid low-thickness regions that is highly robust, easy to control, and exhibits high tolerance with respect to the choice of parameters.

As its main contribution, this work addresses another challenge inherent to VTTO, namely, the formation of blurred structural edges. Blurred edges are a natural consequence of employing any type of blurring filter, i.e., a radius or Helmholtz-type PDE filter \cite{lazarov2011filters}. The aforementioned low-thickness projection methods partially mitigate the issue, forming sharp structural edges, however, only of the thickness $\rho_{\rm low}$.
Sharp structural edges of thickness $\rho > \rho_{\rm low}$ naturally occur in VTTO problems, as seen in the \textit{unfiltered} density field. However, they are entirely eliminated when a blurring filter, necessary for well-posedness of TO, is applied. In standard TO, where black-and-white designs are desired, this issue is resolved by applying the smoothed Heaviside projection which simply pushes each filtered density either towards 0 or 1 (\cite{wang2011projection}). In VTTO, where any density in the range $\rho \in [0,1]$ is equally permitted, this form of Heaviside projection cannot be employed. Therefore, in this work we propose a projection method that restores the original sharpness of structural edges in VTTO problems and enables fully robust VTS designs with clear structural edges of any thickness in the range $\rho \in [\rho_{\rm low},1]$. The proposed projection method utilizes information about the density gradient to target the structural edges specifically while minimally affecting the structure's interior. On top of that, an improved strategy to handle low-thickness regions is proposed which allows a selection of more aggressive continuation parameters and to obtain stable solutions each time. 

This article is organized as follows. Section \ref{sec:vtto} briefly introduces the concept of VTTO in the context of density-based TO. In Section \ref{sec:lt} we propose and test the improved and highly robust approach to eliminate undesired low-thickness regions. The main contribution of this work, that is, the density-gradient-informed (DGI) projection for sharp structural edges, is introduced and tested in Section \ref{sec:dgi}. Final conclusions and outlook are given in Section \ref{sec:concl}.

\section{Variable thickness topology optimization}\label{sec:vtto}

\cite{rossow1973finite} introduced the VTS problem by assigning a thickness variable $\tau_i$ to each finite element. Bounded by the thickness constraints $\tau_i^L \le \tau_i \le \tau_i^U$, the method inherently avoided low thickness regions. However, no topological changes nor formation of intra-domain structural edges were possible. The original VTS problem is, however, closely related to conventional density-based TO, in which the penalization factor is omitted from the SIMP method. By adopting the densities $\rho_e$ as design variables, the VTTO dealing with a volume-constrained compliance minimization problem is formulated as follows

\begin{equation}
	\begin{split}
		\min_{\forall \rho}  \ &: \ \mathcal{F}_{\textup{c}}\left(\rho, \tensor u\right) = \IntSB \tensor u\left(\rho\right) \cdot \tensor{t}_0 \text{\:dA}, \\
		\textup{s.t.}\
		&: \ \mathcal{G}_{\textup{vol}}\left(\rho\right) = \frac{\IntB \rho\left(\tensor{X}\right) \text{\:dV}}{V_0} - \overline{V}_{\rm frac} \leq \ 0, \\
		&: E(\rho_e) = \left[\rho_e\left[1 - \rho_{\rm min}\right] + \rho_{\rm min}\right]E_{0},\\
		&: \ 0 \leq \ \rho_e \leq 1 \ \ \ e = 1,...,N_e,
	\end{split}
	\label{eq:compliance}
\end{equation}
where $\mathcal{F}_{\textup{c}}\left(\rho, \tensor u\right)$ is the compliance objective and $\mathcal{G}_{\textup{vol}}\left(\rho\right)$ is the volume fraction constraint. $E(\rho_e)$ is the interpolated element Young's modulus, for which the element density $\rho_e$ is applied without the penalty exponent $p$ - recall the SIMP formula $E(\rho_e) = \rho_e^p E_0$. The minimum value of $\rho_{\rm min} = 10^{-9}$ is understood as a void, hence permitting the formation of intra-domain void regions. Unlike in conventional TO, intermediate values of $\rho_e \in (0,1)$ are physically meaningful and desired, since they represent the out-of-plane thickness of an extruded structure. Since the penalty exponent $p$ is a restrictive measure, enabling only black-and-white structures, it effectively shrinks the design space. As a consequence, inferior designs in terms of the performance objective are obtained, as shown by \cite{sigmund2016non}, with compliance approximately $30\%$ larger as compared to sheet-like structures. 

\begin{figure}[h!]
	\centering
	\includegraphics[width=84mm]{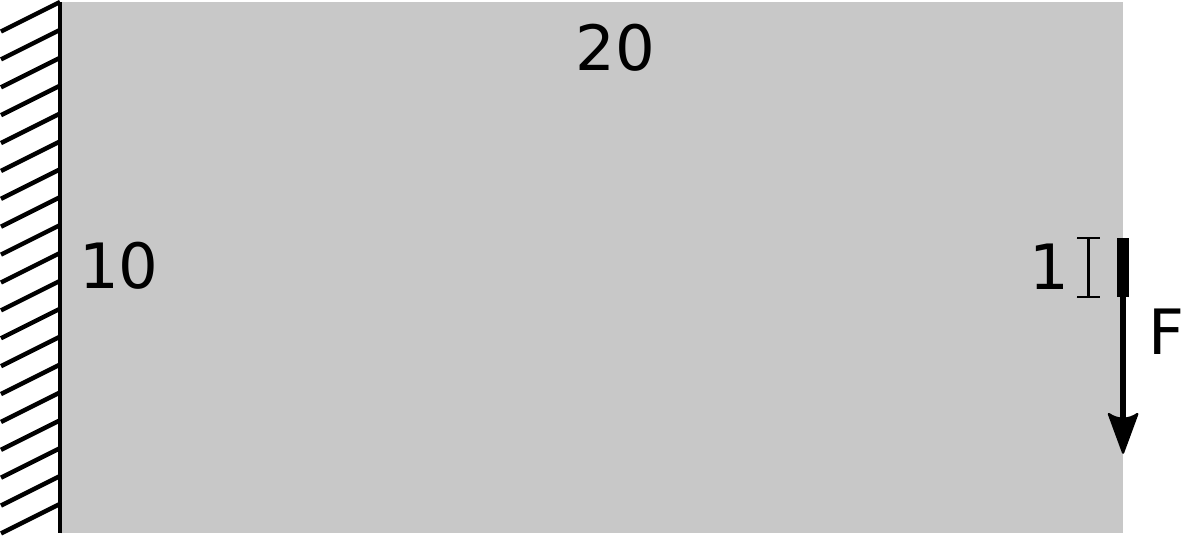}
	\caption{Standard setup for the cantilever benchmark used throughout this work.}
	\label{fig:c-setup}
\end{figure}

For all numerical tests in this work, we use the common setup for the cantilever benchmark, as shown in Fig. \ref{fig:c-setup}.

The parameter setup, common for all examples used throughout this work - unless otherwise specified - is summarized in Table. \ref{tab:setup}.

\begin{table}[h!]
	\centering
	\captionsetup{width=\columnwidth}
	\caption{General parameter setup for the numerical tests.}
	\label{tab:setup}
	\begin{tabular}{cc}
		\hline
		\multicolumn{2}{c}{\textbf{General}} \\
		$\rho_{\rm init}$ & 0.3 \\
		$\overline{V}_{\rm frac}$ & 0.3 \\
		\hline
		\multicolumn{2}{c}{\textbf{Material}} \\
		Type & Linear elastic \\
		E & 1 \\
		$\mu$ & 0.3 \\
		\hline
		\multicolumn{2}{c}{\textbf{Optimizer}} \\
		Type & GOCM \\
		Initial step length & 0.05 \\
		Step length decay & 0.98 \\
		Min. step length & $10^{-4}$ \\
		Design update & exponential \\
		\hline
		\multicolumn{2}{c}{\textbf{Filter}} \\
		Type & PDE \\
		Radius & 0.375 ($1.5h_2$) \\
		\hline
		\multicolumn{2}{c}{\textbf{Stopping criterion}} \\
		Type & $\Delta\rho_{\rm mean}$ \\
		Tolerance & $10^{-4}$ \\
		\hline
		\multicolumn{2}{c}{\textbf{Adaptive mesh}} \\
		Initial ref. level & 2 \\
		Refinenent crit. & $\left\llbracket \rho_e \right\rrbracket_{\max} > 0.05$ \\
		Coarsening crit. & $\left\llbracket \rho_e \right\rrbracket_{\max} < 0.001$ \\
		Max ref. level & 4 \\
		Min ref. level & 0 \\
		Level 0 elem. size & $h_0 = 1$\\
		\hline
	\end{tabular}
\end{table}

The implemented optimizer is the Generalized Optimality Criteria Method (GOCM), as proposed by \cite{kim2021generalized}. In our experience, GOCM deals excellently with both pure compliance minimization problems and multi-constraint problems, i.e., including stress constraints, while maintaining the simplicity level of the original OCM method. The step length is multiplied by the decay parameter in every iteration to aid convergence. Helmholtz-type PDE filter is employed for regularization (\cite{lazarov2011filters}). Since we use adaptive coarsening and refinement, we use the concept of refinement levels, where the coarsest allowable level 0 corresponds to an element size of $h_0 = 1$ and the finest allowable level 4 corresponds to an element size of $h_4 = 1 \cdot 0.5^4 = 0.06125$. The initial mesh level is 2, corresponding to a $80 \times 40$ grid and an element size of $h_2 = 1 \cdot 0.5^2 = 0.25$. For a detailed description of how mesh adaptivity is handled, refer to \cite{stankiewicz2025configurational} and implementation details of the $deal.II$ FEM library \cite{bangerth2007deal, arndt2021deal}. The criterion for mesh adaptation is the density jump $\left\llbracket \rho_e \right\rrbracket_{\max}$. That is, for each element $e$, we find the maximum density difference between the element $e$ and its neighbors, as introduced in \cite{stankiewicz2025novel}. 

\begin{remark}
We emphasize that mesh adaptivity is neither required for the proposed methodology, nor it affects the main findings in this work. While beneficial for computational efficiency, adaptivity introduces implementation complexity. A fixed grid of $320 \times 160$ cells achieves the same resolution for the cantilever benchmark and remains feasible on standard hardware.
\end{remark}

\begin{figure}[h!]
	\centering
	\includegraphics[width=84mm]{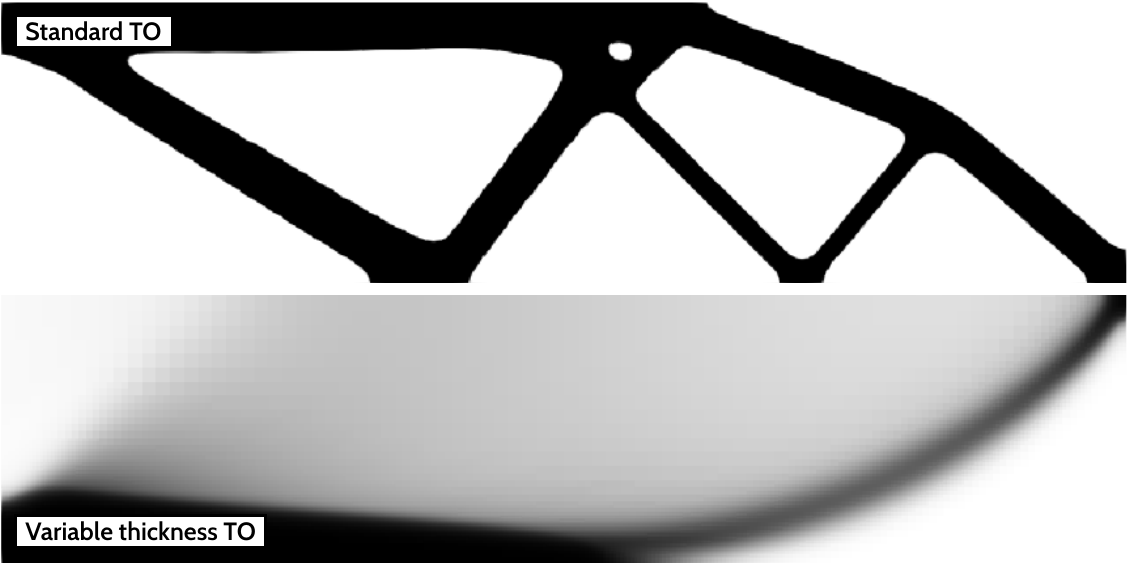}
	\caption{The common benchmark problem - the cantilever - is compared using standard SIMP-based TO and VTTO. The final compliance values are $\mathcal{F}_{\textup{c}} = 7.61 \cdot 10^{-5}$ and $\mathcal{F}_{\textup{c}} = 8.92 \cdot 10^{-5}$, respectively, indicating a $17.2\%$ increase as a result of density penalization.}
	\label{fig:vtto-compare}
\end{figure}

For the stopping criterion, we use the mean density change, accounting for varying element sizes, defined as

\begin{equation}
	\Delta \rho_{\rm mean} = \frac{\sum_e v_e \Delta \rho_e}{\sum_e v_e}.
\end{equation}
where $\Delta \rho_e$ is the density change between optimization iterations for the element $e$ and $v_e$ is the volume of the element $e$. In Fig. \ref{fig:vtto-compare} a comparison between standard and VTTO for the cantilever benchmark is shown. Penalization of intermediate densities results in an around $17.2\%$ increase in the compliance objective.

While VTTO enables superior performance due to an enlarged design space, it faces several issues, like very thin thickness regions and blurred structural edges. These issues are addressed in the following sections.

\section{Generalized treatment of low thickness regions}\label{sec:lt}

\begin{figure}[h!]
	\centering
	\includegraphics[width=84mm]{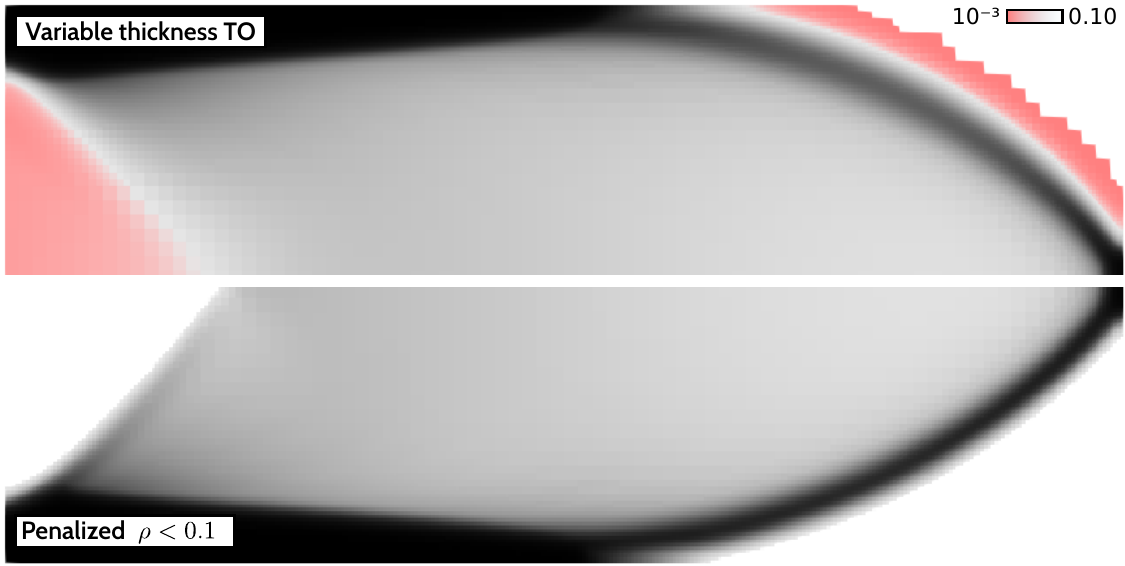}
	\caption{Penalization of low-thickness densities (here $\rho < 0.1$) helps generate manufacturable structures with a negligible cost to the performance objective. In this case, the penalization caused merely a $0.26\%$ increase in compliance.}
	\label{fig:vtto-lt}
\end{figure}

VTTO inherently produces very thin sheets, which are not only difficult to interpret and manufacture, but also do not contribute significantly to the objective value, as pointed out by \cite{giele2021approaches}. Hence, penalization of low-thickness densities is necessary to obtain meaningful structures, see Fig. \ref{fig:vtto-lt}.

We classify density values in the range $\rho \in (0, \rho_{\rm low})$ as unacceptably thin sheets, where $\rho_{\rm low}$ is a critical value below which we deem the resulting thickness undesirable. There are several approaches to mitigate low-thickness regions in VTTO, either focusing on introducing a selective, SIMP-based penalization (\cite{endress2023designing}), projection-based approaches in the regularization step (\cite{groen2018homogenization, larsen2018optimal}) or utilizing an additional design field (\cite{giele2021approaches}). 

In the following, we combine the SIMP-based and the projection-based approaches, closely following the logic of standard TO, where sharp black-and-white structures are obtained using both penalization and projection. This is also motivated by findings that using only either the penalization-based or projection-based approaches either does not strictly eliminate low-thickness densities or does not converge to meaningful structures, exhibiting high sensitivity to continuation schemes. Our numerical studies have shown that using a combination of SIMP-based and projection-based approaches leads to stable and effective suppression of low-thickness regions, even with aggressive parameter continuation strategies.

\subsection{SIMP-based suppression}\label{sec:simp}

As the first component of the strategy for treating low-thickness regions, we consider the penalization-based approach as proposed by \cite{endress2023designing}:

\begin{equation}
	\rho_e^{\rm VT} =
	\begin{cases}
		\rho_e,& \text{if }  \rho_e \geq \rho_{\rm low}\\
		\left(\frac{\rho_e}{\rho_{\rm low}}\right)^{p}\rho_{\rm low},& \text{else}.
	\end{cases}
	\label{eq:pen}
\end{equation}
where $\rho_e^{\rm VT}$ is the selectively penalized density, with the penalty exponent $p$ as in standard TO. The per-element Young's modulus is then computed as

\begin{equation}
	E(\rho_e) = \left[\rho_e^{\rm VT}\left[1 - \rho_{\rm min}\right] + \rho_{\rm min}\right]E_{0}, 
	\label{eq:simp}
\end{equation}

The penalty factor is chosen to be $p = 3$. In order to allow for a full design freedom in the initial iterations of optimization, a simple continuation approach is implemented

\begin{enumerate}
	\item Start with $p = p_{\rm init} = 1$
	\item Perform in every iteration:
	\begin{equation}
		p^{I+1} = \min (c_p p^I, p_{\rm max})
	\end{equation}
	with $p_{\rm max} = 3$.
\end{enumerate}

\begin{figure}[h!]
	\centering
	\includegraphics[width=84mm]{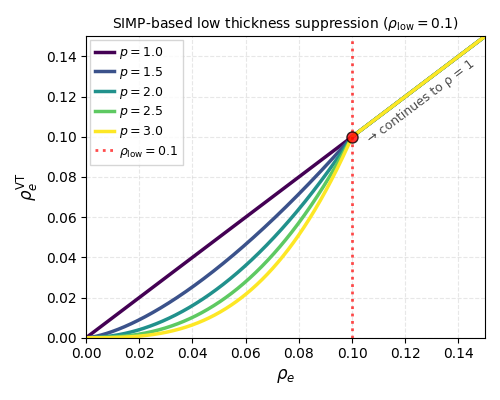}
	\caption{SIMP-based penalization for the suppression of low-thickness regions for $\rho_{\rm low} = 0.1$, including the continuation for the penalty exponent: $p = 1 \to 3$.}
	\label{fig:plot-pen}
\end{figure}

The evolution of the penalized density $\rho_e^{\rm VT}$ during the continuation is visualized in Fig. \ref{fig:plot-pen}.

This penalization approach contains a discontinuity at $\rho_e = \rho_{\rm low}$, hence, the derivative of Eq. \ref{eq:pen} is not available at $\rho_{\rm low}$. Consequently, this approach is mathematically inconsistent regarding gradient computation. However, considering that the numerical treatment involves floating-point arithmetic, the elemental densities never strictly fulfill the condition $\rho_e = \rho_{\rm low}$. Hence, for any numerical value of the density $\rho_e$, an inequality condition is sufficient to determine whether the sensitivity should be computed for $\rho_e < \rho_{\rm low}$ or otherwise. A continuous version of Eq. \ref{eq:pen} can be formulated using, for instance, a sigmoid-type switching function. However, to keep the formulation simple, the current, discontinuous version of Eq. \ref{eq:pen} is chosen.

\subsection{Updated projection-based suppression}\label{sec:lt-proj}

\begin{figure}[h!]
	\centering
	\includegraphics[width=84mm]{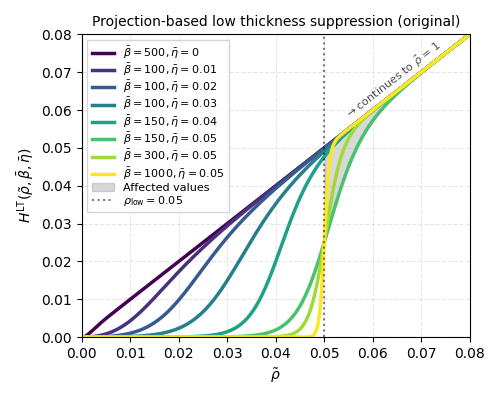}
	\caption{Projection-based suppression function for low-thickness regions as proposed by \cite{larsen2018optimal}, plotted for each pair of parameters $\{\bar\beta,\bar\eta\}$ used in the continuation scheme. The grey region marks the impact of the projection on densities $\rho > \rho_{\rm low}$.}
	\label{fig:plot-lt-old}
\end{figure}

The second component of the strategy to treat low-thickness regions involves a more common approach, namely, a projection function in the regularization step. The projection step employs a form of sigmoid-type smoothed Heaviside projection, as first introduced by \cite{wang2011projection}. Adapted functions to target only low-thickness regions were proposed in \cite{groen2018homogenization, larsen2018optimal}. These approaches, however, use a continuation strategy for both the sharpness $\bar\beta$ and threshold $\bar\eta$ using  predetermined pairs of $\{\bar\beta,\bar\eta\}$ values. The version of the projection of \cite{larsen2018optimal} is shown in Fig. \ref{fig:plot-lt-old}. The evolution of the $\bar\beta$ parameter is non-monotonic, making it difficult to adjust for various problems. For different low-thickness thresholds, a new set of predetermined pairs of $\{\bar\beta,\bar\eta\}$ needs to be defined.

In order to tackle the aforementioned challenges, we adapt the projection function to fulfill the following criteria: (i) the continuation strategy is based purely on changing the projection sharpness $\bar\beta$; (ii) the continuation strategy is controlled by a single parameter in the manner $\bar\beta^{I+1} = \bar c_\beta \bar\beta^I$; (iii) the parameter choice is independent of the low-thickness threshold $\rho_{\rm low}$; (iv) the values $\rho > \rho_{\rm low}$ should be minimally affected. The criteria (i) and (ii) are formulated in analogy to the smoothed Heaviside projection used for standard TO (\cite{wang2011projection}). By employing analogous parameter selection and continuation strategies as in standard TO, we point to several practical benefits such as intuitive parameter selection and a clean, consistent code implementation in case both standard TO and VTTO coexist in the same code. 

Thus, we employ an approach that utilizes two basic functions: the $\bar\beta$-penalized density $\tilde\rho^{\bar\beta}$ and the identity $\tilde\rho$ which are interpolated using a switching function $S(\tilde\rho,\bar\beta,\rho_{\rm low})$ in the following manner:

\begin{equation}
	H^{\rm LT}(\tilde\rho,\bar\beta) = [1 - S(\tilde\rho,\bar\beta,\rho_{\rm low})] \tilde\rho^{\bar\beta} + S(\tilde\rho,\bar\beta,\rho_{\rm low}) \tilde\rho
\end{equation}

where the switching function $S(\tilde\rho,\bar\beta,\rho_{\rm low})$ is a sigmoid function given by

\begin{equation}
	S(\rho,\beta,\eta) = \frac{1}{2} \left[1 + \tanh\left(\beta \left[ \frac{\rho}{\eta} - \eta^{1/\beta} \right]\right)\right].
\end{equation}

\begin{figure}[h!]
	\centering
	\includegraphics[width=84mm]{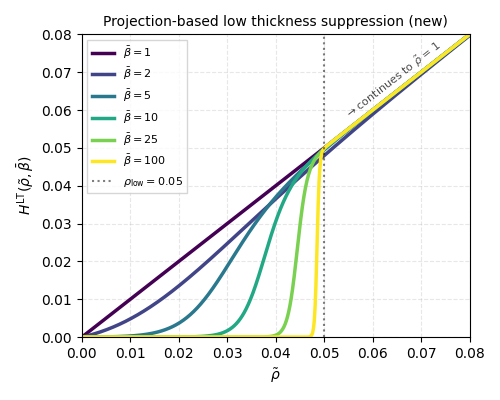}
	\caption{Updated version of the projection-based suppression function for low-thickness regions. The new version exploits only the $\bar\beta$ parameter for the continuation in a robust, $\rho_{\rm low}$-independent way. Densities $\rho > \rho_{\rm low}$ are minimally affected.}
	\label{fig:plot-lt-new}
\end{figure}

The switching function $S(\rho,\beta,\eta)$ is similar to the known smoothed Heaviside projection introduced by \cite{wang2011projection} with the difference that $S(\rho,\beta,\eta)$ is a rather standard sigmoid-type function with a projection range of $\bar\rho \in (0,1)$, as opposed to one exactly passing through points $\{0,0\}$ and $\{1,1\}$. The characteristic modification to the sigmoid function is the threshold shifting part: $\left[ \rho/\eta - \eta^{1/\beta} \right]$, where incorporation of the sharpness parameter $\beta$ enables threshold shifting behavior similar to the approach of \cite{larsen2018optimal}, without the need to explicitly define the continuation values for $\eta$. In Fig. \ref{fig:plot-lt-new}, the updated projection function is plotted for a selection of $\beta$ values.

Starting with $\bar\beta = 1$, the projection $H^{\rm LT}(\tilde\rho,\bar\beta)$ is an identity function $H^{\rm LT}(\tilde\rho,1) = \tilde\rho$, since both basic functions $\tilde\rho^1$ and $\tilde\rho$ are identity functions. An increase in $\bar\beta$ gradually penalizes $\tilde\rho^{\bar\beta}$ while at the same time the switching function $S(\tilde\rho,\bar\beta,\rho_{\rm low})$ becomes sharper. Most notably, densities $\tilde\rho > \rho_{\rm low}$ are only marginally affected while the behavior for $\tilde\rho < \rho_{\rm low}$ is similar to the original projection in Fig. \ref{fig:plot-lt-old}, with only $\beta$ being updated. As for the SIMP-based suppression, we employ a continuation strategy in an analogous manner:

\begin{enumerate}
	\item Start with $\bar\beta = \bar\beta_{\rm init} = 1$
	\item Perform in every iteration:
	\begin{equation}
		\bar\beta^{I+1} = \min (\bar c_\beta \bar\beta^I, \bar\beta_{\rm max})
	\end{equation}
\end{enumerate}

\subsection{Combined suppression}

Ultimately, we aim to use the SIMP-based and the projection-based approaches to suppress low-thickness regions together. In order to arrange for this, a joint continuation scheme is used in a manner analogous to multi-thickness topology optimization \cite{stankiewicz2025novel}, where the continuation of $p$ and $\bar\beta$ is arranged sequentially. This order of sequence has worked very well for various increase rates in multi-thickness TO and, similarly, works well in the proposed low-thickness suppression approach in this work. The continuation scheme is arranged as follows:

\begin{enumerate}
	\item Start with $p = p_{\rm init} = 1$ and $\bar\beta = \bar\beta_{\rm init} = 1$
	\item Perform in every iteration:
	\begin{equation}
		p^{I+1} = \min (c_p p^I, p_{\rm max})
	\end{equation}
	where $p_{\rm max} = 3$.
	\item Once $p = p_{\rm max} = 3$, perform in every iteration:
	\begin{equation}
		\bar\beta^{I+1} = \min (\bar c_\beta \bar\beta^I, \bar\beta_{\rm max})
	\end{equation}
\end{enumerate}

Although sequential suppression is preferred due to higher stability during optimization, our additional studies have shown that a simultaneous continuation of $p$ and $\bar\beta$ can arrive at satisfactory designs as well. However, more attention is required for the appropriate selection of the increase rates. Ultimately, in combination with the DGI projection, which is introduced in Section \ref{sec:dgi}, the sequential continuation approach is strongly advised.

\subsection{Numerical example}

\begin{figure*}[h!]
	\centering
	\includegraphics[width=174mm]{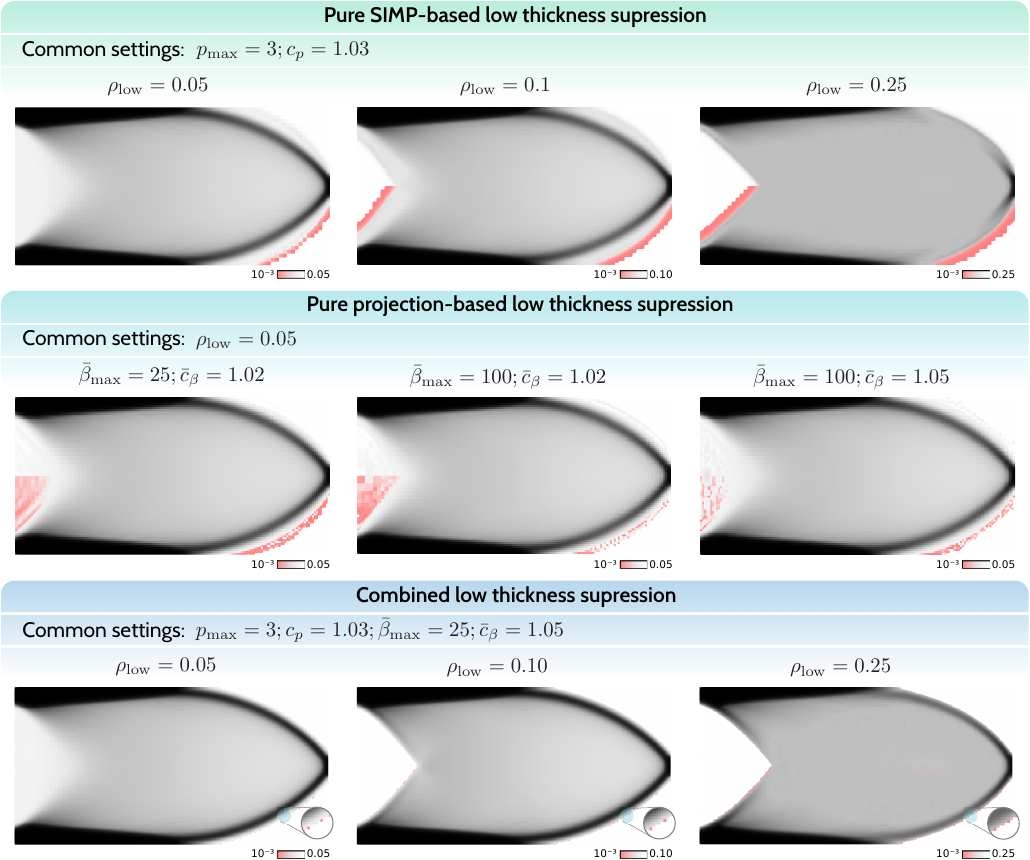}
	\caption{Numerical tests for the studied low-thickness suppression approaches. From top to bottom: pure SIMP-based, pure projection-based, and combined suppression. Undesired low-thickness densities in the range $\rho \in (0.001, \rho_{\rm low})$ are highlighted using a red-to-transparent scale. Using the combined approach, the amount of low thickness densities is negligible.}
	\label{fig:lt-study}
\end{figure*}

In the following, we test the effectiveness of the introduced strategies involving the SIMP-based and the projection-based low-thickness suppression. First, we are interested in how the SIMP-based and projection-based approaches perform alone using our continuation approach. In particular, for the projection-based approach, the proposed continuation strategy from Section \ref{sec:lt-proj}, using the increase rates of our choice, is more aggressive compared to the reference method from Fig. \ref{fig:plot-lt-old}, for which an interval of 50 iterations between each continuation step is chosen. Ultimately, the combined suppression approach is tested and compared against the pure SIMP-based and projection-based cases. The basis for assessment is the presence of undesired low-thickness densities in the range $\rho \in (0.001, \rho_{\rm low})$. The values $\rho < 0.001$ are considered acceptable void. For visualization, we employ a reversed red-to-transparent scale for the $\rho \in (0.001, \rho_{\rm low})$ range to draw attention to the most undesired densities, namely the values close to the void threshold. The red-to-transparent scale is only applied to the bottom half of the structures, exploiting the symmetry of the designs. In Fig. \ref{fig:lt-study} the resulting structures are shown.

The first row in Fig. \ref{fig:lt-study} depicts the resulting structures for the pure SIMP-based suppression for varying low-thickness thresholds $\rho_{\rm low} = \{0.05,0.1,0.25\}$. The maximum penalty value $p_{\max}$ and the increase rate $c_p$ are kept constant, as their variation revealed no significant design changes. SIMP-based suppression effectively forms structures that respect the low-thickness threshold, similarly to standard TO, due to penalized stiffness and, therefore, undesirably high compliance in the low-thickness regime. However, due to the blurring filter, low-thickness transition regions of width proportional to the filter radius occur between solid and void. This indicates the need for a projection-based treatment after applying the filter.

In the second row, the resulting structures for the pure projection-based approach are shown. This approach, using the function from Fig. \ref{fig:plot-lt-old}, is successfully employed in literature, however, using a less aggressive continuation strategy that spans approximately 300-400 optimization iterations. In this work, we test a more aggressive continuation scheme that usually spans over 100-150 iterations, which naturally risks decreased optimization stability. The results revealed just that. Even for the low-thickness threshold of $\rho_{\rm low} = 0.05$, the final designs contain large regions with undesired low-thickness densities. Hence, for the pure projection-based study, we focused on the influence of the final projection sharpness $\bar\beta_{\max}$ and the increase rate $\bar c_\beta$ to find out whether an improvement can be achieved. Unfortunately, variations of these parameters did not yield significant changes. Testing the original approach from Fig. \ref{fig:plot-lt-old}, but using our continuation setup, revealed similar behavior. The large presence of undesired low-thickness densities can be justified by the fact that low-thickness densities are actually still desired by the optimizer. In the physical sense, they do not come at an additional cost in terms of compliance as in the case of the SIMP-based approach. Moreover, the projection does not approach the near-perfect sharpness of the Heaviside function, as would be the case for $\bar\beta_{\max} \to \infty$. Therefore, there is a small range of densities for which low-thickness densities are possible, which pushes the optimizer to find that range if it minimizes the objective function. The higher the projection sharpness, the more noisy the low-thickness regions, while still being largely present.

The observations from the pure SIMP-based and projection-based approaches naturally suggest a combined approach to fully eliminate low-thickness regions in a stable manner. While the SIMP-based approach informs the optimizer that low-thickness densities are undesired to minimize the objective, the projection-based approach can assist with eliminating the filter-induced low-thickness solid-to-void transitions. The last row in Fig. \ref{fig:lt-study} shows the resulting structures employing the combined approach. We chose to showcase the designs for various low-thickness thresholds, as the influence of various increase rates, penalties, and sharpness parameters was minimal. The resulting designs are satisfactory with an almost complete absence of low-thickness densities. By using relatively aggressive continuation parameters, the robustness of the combined approach was proven. Moreover, the projection sharpness of $\bar\beta_{\max} = 25$ was sufficient for effective elimination of the low-thickness densities in the solid-to-void transition.

\subsection{The remaining challenge: low thickness edges} 

\begin{figure*}[h!]
	\centering
	\includegraphics[width=174mm]{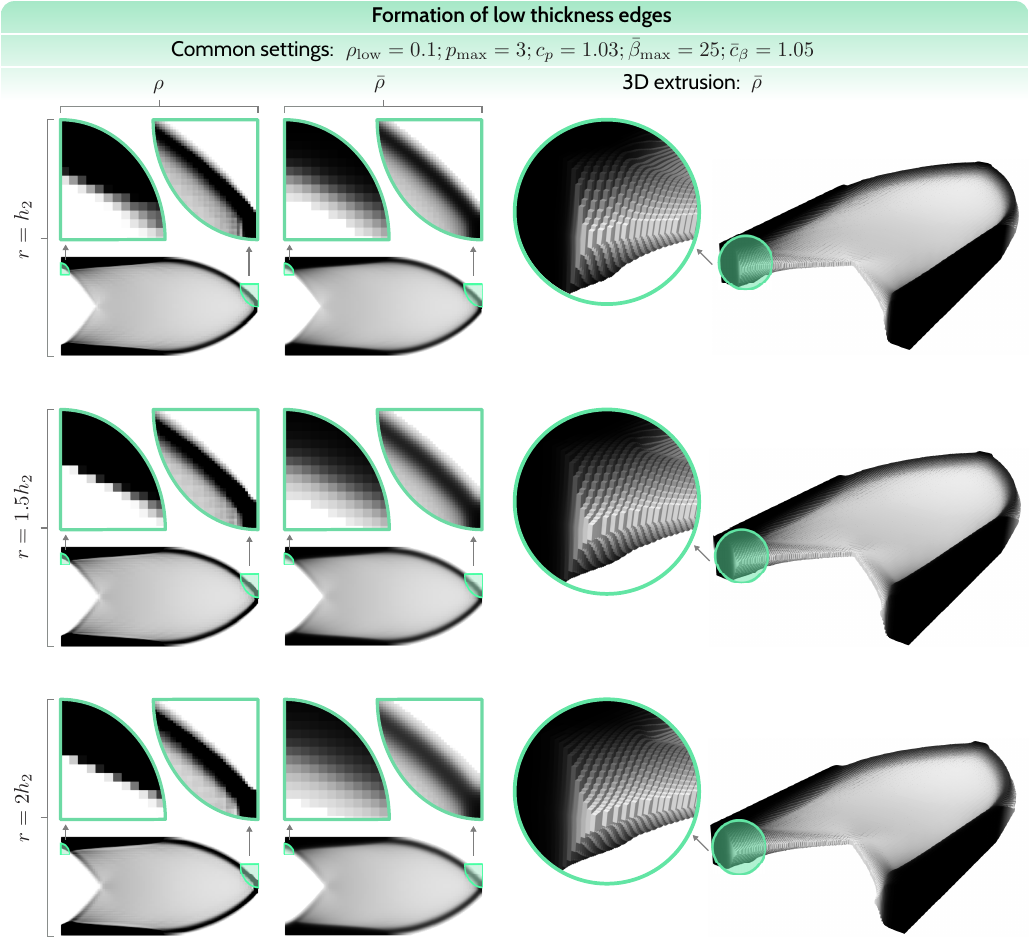}
	\caption{Influence of the common regularization scheme, including the blurring filter and the low-thickness projection, on the formation of structural edges. Three cases with various filter radii $r \in \{h_2, 1.5h_2, 2h_2\}$ are compared.}
	\label{fig:lt-edges}
\end{figure*}

The combined low-thickness suppression approach has proven to be a reliable solution to eliminate low-thickness regions. However, a characteristic occurrence in such designs is the formation of structural edges with the exact thickness of the threshold parameter $\rho_{\rm low}$. This is a natural consequence of the combined regularization using a blurring filter and low-thickness projection, necessary to deal with checkerboards and mesh dependencies. Such low-thickness edges are closely inspected in Fig. \ref{fig:lt-edges}.

To understand the influence of the PDE filter, in particular the choice of the filtering radius, we compare three cases: $r \in \{h_2, 1.5h_2, 2h_2\}$. Naturally, the optimizer tries to form structural edges of thicknesses also above $\rho_{\rm low}$, as can be seen in the raw (unfiltered) density field, see the first column in Fig. \ref{fig:lt-edges}. We determined two critical locations where the influence of density regularization on the formation of structural edges is clearly visible. Close-up views of these regions for both the raw densities $\rho$ and regularized densities $\bar\rho$ reveal a substantial difference in the sharpness of the structural edges. The optimizer attempts to form a sharp structural edge of thickness $\rho = 1$ near the Dirichlet boundary condition. Interestingly, as the filter radius increases, the structural edge in the $\rho$ field gets sharper, suggesting that the optimizer more strongly prevents the formation of a blurred low-thickness edge.

A better intuition for the low-thickness edges can be obtained with the help of 3D visualization by a simple extrusion of the finite elements in the out-of-plane direction, proportionally to their thicknesses $\bar\rho$. The structural edges are clearly formed by density values of $~ \rho_{\rm low}$, with a decreasing density slope for an increasing filter radius.

Clearly, low-thickness edges are a filtering artifact. The current regularization strategy prevents the formation of natural structural edges of any thickness. While this occurrence is not a major hindrance for simple compliance minimization problems, it is highly undesired for problems that contain a singularity-like feature, for instance, the L-beam. The reason for this is the accumulation of stresses in the vicinity of the singularity, which will be amplified if the structural edges can only be of low-thickness $\rho_{\rm low}$. While L-beam type problems and stress-constrained optimization are beyond the scope of this paper, it is important to consider the bigger picture of VTTO when addressing the formation of structural edges.

\section{Density-gradient-informed projection for sharp edges}\label{sec:dgi}

The formation of blurred, low-thickness structural edges in VTTO is inherent to the common regularization scheme involving a blurring filter, like the Helmholtz-type PDE filter.

Thus, to address this issue, we propose a projection method to deblur (resharpen) structural edges as an additional step in the regularization workflow. The projection needs to fulfill the following requirements: (i) restore the sharpness of structural edges as in the raw density field $\rho$; (ii) target structural edges and minimally affect the interior of the structure; (iii) use an intuitive $\beta$-like parameter to control the intensity of deblurring. A targeted effect on the structural edges and not the interior of the structure is essential to preserve the "variable thickness" character of the method. 

\begin{figure}[tb]
	\centering
	\includegraphics[width=84mm]{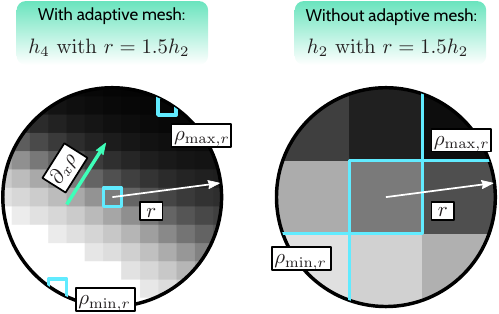}
	\caption{Neighborhood $N_r(e)$ and the essential quantities necessary for DGI projection, shown for cases with and without adaptive meshing.}
	\label{fig:dgi-terms}
\end{figure}

In the following, we first introduce the terms used in the proposed projection. The neighborhood of an element $e$ encompasses elements within the distance of the filter radius $r$:

\begin{equation}
	N_r(e) = \{ e' \in \Omega_e \mid \| \mathbf{x}_{e} - \mathbf{x}_{e'} \| \le r \},
\end{equation}
where $e'$ indicates the elements that are neighbors of $e$ and $\mathbf{x}_{e}$ is the center of element $e$. As the selected filter radius for our numerical tests of $r = 1.5h_2$ is based on the initial mesh size of $h_2 = 0.25$, and we utilize adaptive meshing, in Fig. \ref{fig:dgi-terms} we show the neighborhood $N_r(e)$ for a case with adaptive meshing of up to $h_4$ and with a uniform mesh, using $h = h_2$. For uniform meshes, it is recommended to use a minimum neighborhood radius of $r_{\min} = 1.5h$ to ensure the immediate neighborhood of $e$ exists, even when the filter radius is smaller. For $r < h$, the neighborhood $N_r(e)$ contains only the element $e$ itself, causing the proposed deblurring method to have no effect.
Within the neighborhood $N_r(e)$ we find the minimum and maximum density

\begin{equation}
	\rho_{\min,r} (e) = \min_{e' \in N_r(e)} \{ \rho_{e'} \},
\end{equation}

\begin{equation}
	\rho_{\max,r} (e) = \max_{e' \in N_r(e)} \{ \rho_{e'} \}.
\end{equation}

We define the density variation within the neighborhood $N_r(e)$ as the difference between the maximal $\rho_{\max,r} (e)$ and the minimal $\rho_{\min,r} (e)$ densities

\begin{equation}
	d_r(e) = \rho_{\max,r} (e) - \rho_{\min,r} (e) \approx \partial_x \rho (e) \cdot 2r
\end{equation}
which is essentially a finite difference approximation of the density gradient $\partial_x \rho (e)$ multiplied by the neighborhood diameter $2r$, assuming $\rho_{\max,r} (e)$ and $\rho_{\min,r} (e)$ are found close to the boundary of $N_r(e)$. Finally, we define the median density as

\begin{equation}
	\rho_{{\rm mid},r} (e) = \rho_{\min,r} (e) + 0.5 d_r(e).
\end{equation}

Having all the necessary quantities at hand, we propose the deblurring projection function as a \textit{scaled} and \textit{shifted} smoothed Heaviside projection

\begin{equation}
	\begin{split}
		H^{\rm DGI}_e (\tilde\rho_e^{\rm local}, \hat\beta_d(e), \hat\eta(e)) &=\\ d_r(e)H(\tilde\rho_e^{\rm local}, &\hat\beta_d(e), \hat\eta(e)) + \rho_{\min,r}(e)
	\end{split}
	\label{eq:dgi}
\end{equation}
where the smoothed Heaviside projection takes the standard form introduced by \cite{wang2011projection}

\begin{equation}
	H(\rho,\beta,\eta) = \frac{\tanh\left(\beta \eta\right) + \tanh\left(\beta \left[ \rho - \eta \right] \right)}{\tanh\left( \beta \eta \right) + \tanh \left( \beta \left[ 1 - \eta \right] \right)}.
	\label{eq:heaviside}
\end{equation}

The first essential modification to the standard form of the projection is scaling the values by the density difference $d_r(e)$ and shifting the values upwards by the minimum density $\rho_{\min,r} (e)$. This way, we restrict the projection to operate within the range $\rho_{\min,r} (e) \leq \hat\rho_e \leq \rho_{\max,r} (e)$. At this point, it is appropriate to elaborate that we associate the 'hat' accent with quantities specific to the deblurring projection, e.g. the deblurred density $\hat\rho_e$. 

\begin{figure*}[bt]
	\centering
	\includegraphics[width=174mm]{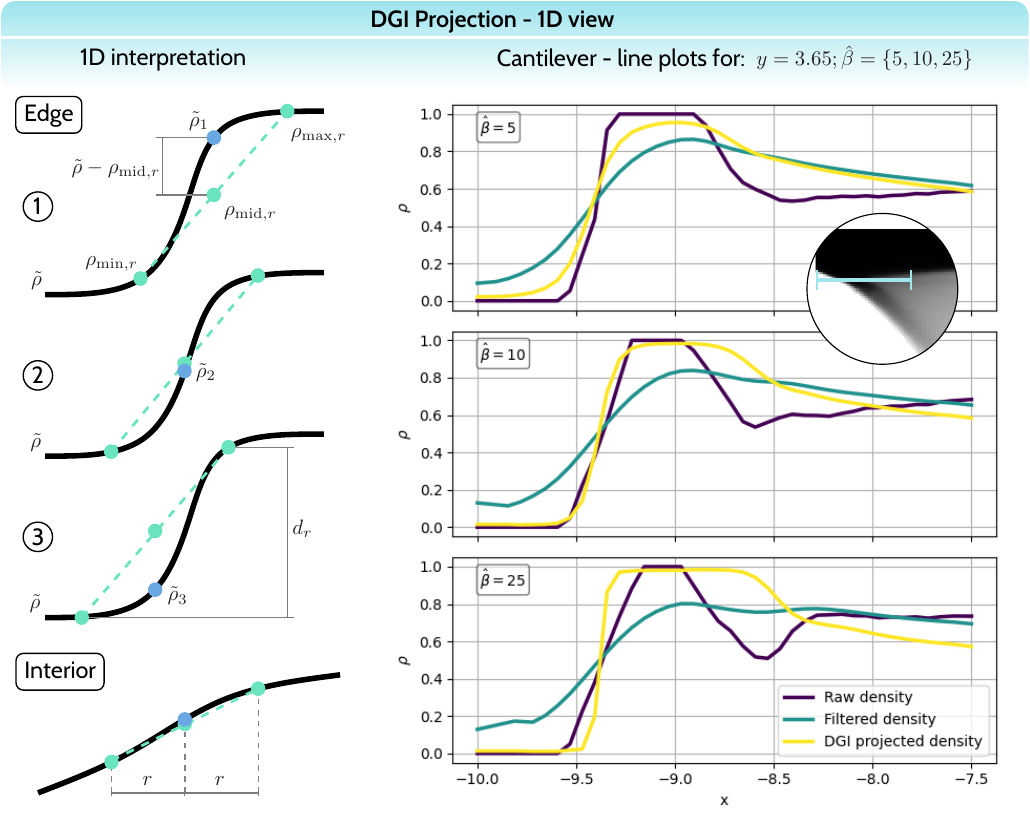}
	\caption{1D view of DGI projection. On the left side, a sketch of a blurred edge is depicted, for which three scenarios are considered, and a representation of the structure's "interior". Based on the difference $\tilde\rho - \rho_{{\rm mid},r}$, the following behavior is expected: In scenario 1), the density $\hat \rho_1$ will be projected upwards; in scenario 2), $\hat \rho_2$ will be barely affected; in scenario 3), $\hat \rho_3$ will be projected downwards; and in the structure's "interior", the projection is expected to be negligible. On the right side, line plots along a selected portion of the cantilever are depicted, for which the raw density $\rho$, the filtered density $\tilde \rho$, and the DGI projected density $\hat \rho$ are plotted.}
	\label{fig:lineplot}
\end{figure*}

The second essential modification is the adjusted parameters passed to the smoothed Heaviside projection in Eq. \ref{eq:dgi}. The passed density is scaled and shifted 

\begin{equation}
	\tilde\rho_e^{\rm local} = \frac{\tilde\rho_e-\rho_{\min,r} (e)}{d_r(e)}
\end{equation}
such that the projection works within the local range $\left[\rho_{\min,r} (e), \rho_{\max,r} (e)\right]$. As the projection threshold, we use the median density $\rho_{{\rm mid},r} (e)$. Similarly, the projection threshold needs to be scaled and shifted for consistency with $\tilde\rho_e^{\rm local}$

\begin{equation}
	\hat\eta (e) = \frac{\rho_{{\rm mid},r}(e)-\rho_{\min,r} (e)}{d_r(e)} = 0.5
\end{equation}
ultimately reducing to $\hat\eta (e) = 0.5$. The key adjustment that defines the character of the DGI projection is applied to the projection sharpness parameter. Namely, we scale the projection sharpness for each element by the associated density difference

\begin{equation}
	\hat\beta_d (e) = \hat\beta d_r(e),
	\label{eq:beta}
\end{equation}
hence the nomenclature density-gradient-informed (DGI) projection. The choice of this scaling is motivated by the observation that the strength of the effect that a blurring filter has is proportional to the gradient of the filtered density field. For instance, consider two locations, for which the raw density field resembles a sharp edge and a relatively smooth structure interior (see Fig. \ref{fig:lineplot}, left side for a demonstrative 1D sketch of an edge and the interior). Application of a blurring filter will smooth out both regions, however preserving a steeper slope at the location of the edge. Hence, in order to specifically target structural edges, the information about this slope is utilized to define the strength of the DGI projection.

In the following, to further clarify the logic behind DGI projection, in Fig. \ref{fig:lineplot}, left side, we consider three different scenarios for determining DGI projection behavior for an edge. The black line representing blurred density $\tilde \rho$ forms an S-shaped curve. Naturally, the bottom half of this curve is locally convex and the top half is locally concave. This property is exposed by comparing the three chosen density values $\tilde \rho$ with the median densities $\rho_{{\rm mid},r}$.\footnote{Strictly speaking, convexity can only be verified by considering all points along the straight line formed by $\rho_{{\rm min},r}$ and $\rho_{{\rm max},r}$, which clearly is not the case here. However, by splitting the curve in Fig. \ref{fig:lineplot} at the inflection point, we obtain fully convex and concave sections. Nevertheless, our goal is to point out a trend which relates the projection direction with local curvature characteristics of the blurred density $\tilde \rho$.} Whether the difference $\tilde\rho - \rho_{{\rm mid},r}$ is positive or negative, the density will be projected upwards or downwards, respectively. Moreover, the larger this difference, the stronger the projection, as known from the characteristics of the smoothed Heaviside projection, since the median density $\rho_{{\rm mid},r}$ contributes to the projection threshold $\hat\eta (e)$.

Next, before the analysis of the complete 2D numerical studies, we explore the effects of DGI projection by extracting density plots along a chosen line for a cantilever case, as shown on the right side of Fig. \ref{fig:lineplot}. At this stage, we do not aim to draw attention to the complete cantilever design, but rather purely focus on the extracted 1D plots. By first understanding the behavior of DGI projection in a 1D regime, we can assess the method better when viewed on a complete 2D structure. The chosen section for the 1D plot is shown in a close-up view below the first plot in Fig. \ref{fig:lineplot}. This location is the most representative as it runs along a structural edge that ideally would form a full-thickness sharp edge, and further along a portion of the structure's interior with intermediate density values. Note that the 1D plots do not represent an actual pure 1D study, since neighbor sets $N_r(e)$ for each element also consider cells up to $r$ distance away in the perpendicular direction to the plotting line.

The 1D plots in Fig. \ref{fig:lineplot} depict three cases with varying projection sharpness $\hat \beta$, see Eq. \eqref{eq:beta}. For each case, three fields were plotted: raw density $\rho$, filtered density $\tilde \rho$ and DGI projected density $\hat \rho$. Note that the raw density forms a relatively sharp edge of full thickness - the raw density steeply transitions from $\rho = 0$ to $\rho = 1$. The filtering process naturally smooths out the edge and even decreases its thickness, but guarantees a more regular density distribution within the structure's interior. Finally, the DGI projection reverses the edge smoothing, each time to a different degree depending on the selected sharpness parameter $\hat \beta$. For $\hat \beta = 5$, the edge sharpness is only partially retrieved, while the structure's interior retains almost the same density distribution. By increasing the sharpness parameter to $\hat \beta = 10$ we arrive at a sweet spot, for which the original edge sharpness is closely matched, while the interior is insignificantly affected. For $\hat \beta = 25$ we obtain an over-sharpened edge with a more notable alteration of the structure's interior. At this point, it is essential to note that while $\hat \beta = 10$ is optimal in this study, the sharpness choice is entirely dependent on the filtering radius for the blurring filter, since DGI projection is basically reversing its effect. Nevertheless, the 1D plots in Fig. \ref{fig:lineplot} should provide a fair initial intuition on how DGI projection works.

\subsection{On the desired edge sharpness}\label{sec:sharpness}

By observing the plots in Fig. \ref{fig:lineplot}, a question might arise as to why we should not aim at obtaining perfectly sharp structural edges, that is, a direct transition from $\hat\rho = 0$ to $\hat\rho = 1$ for adjacent cells. Ideal sharp edges are essentially why the level-set method and node-based shape optimization excel at fine tuning designs to meet certain boundary-sensitive requirements, e.g. curvature control \citep{stankiewicz2022geometrically}. This is possible because the edges are modeled \textit{explicitly}. By default, density-based topology optimization considers a domain composed of a structured mesh. The cells are perfect squares, rendering any structural edges in a \textit{staircase} pattern. Assuming the structural edges are formed as "ideally" sharp, i.e. a direct black-to-white transition, the edge would appear as a pure staircase. Pure staircase edges are not desirable for two reasons. 

First, considering physical implications, perfectly sharp or stepwise edges introduce artificial stress concentrations at element corners, which do not correspond to realistic material behavior. In practice, manufactured structures inevitably exhibit rounded or smoothed edges, and thus the mechanical response of a staircase geometry does not reflect that of a realizable design.

\begin{figure}[tb]
	\centering
	\includegraphics[width=84mm]{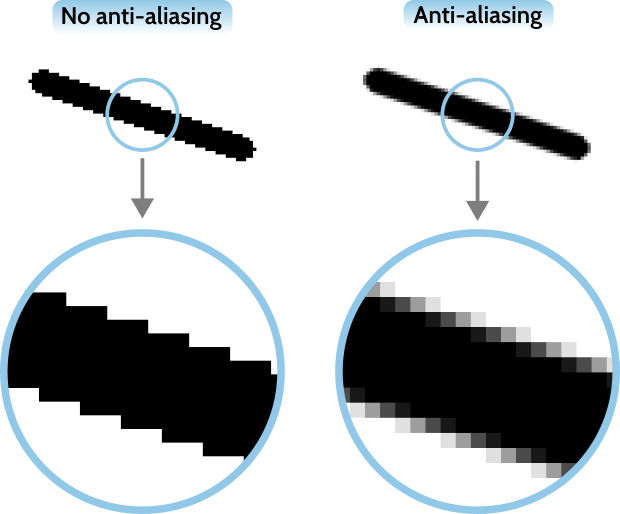}
	\caption{The ideal contour of a pixelated object involves a small amount (pixel-width) of blur at the edges. In computer graphics, this is ensured by applying an anti-aliasing filter, which eliminates the jagged character of edges.}
	\label{fig:anti-aliasing}
\end{figure}

Second, considering a representational viewpoint, a pure staircase edge in fact appears less "sharp" than an edge containing some amount of intermediate densities "filling" the staircase corners. The staircase effect is analogous to aliasing artifacts in digital images, where coarse pixelation leads to jagged contours. In computer graphics, this issue is mitigated through anti-aliasing, which smooths transitions to approximate the underlying continuous form, see Fig. \ref{fig:anti-aliasing} for an example. Maintaining "anti-aliased" structural edges not only improves their visual aspect, but also enables more efficient post-processing tools, rendering more accurate explicit models for manufacturing.

Thus, when deciding on the sharpness parameter $\hat \beta$, we should take these factors into consideration. Ideally, when deblurring structural edges, we should aim at maintaining a cell-width transition zone, which would mitigate the formation of fully jagged edges.

\subsection{Complete workflow}\label{sec:workflow}

\begin{figure}[tb]
	\centering
	\includegraphics[width=84mm]{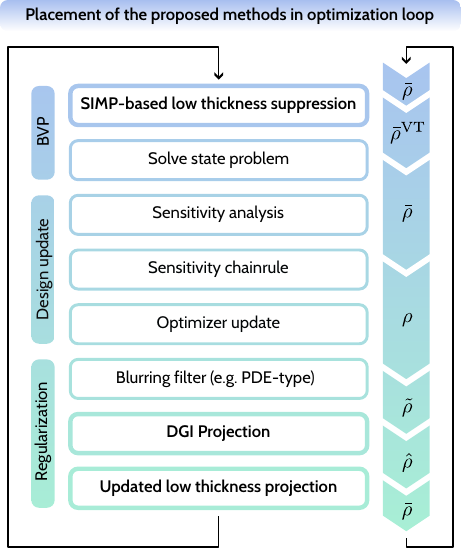}
	\caption{A schematic representation of the optimization loop focusing on the placement of the methods introduced in this work. The SIMP-based low-thickness suppression is executed within the tangent matrix assembly for the state problem. The DGI projection directly follows the filtering (here: PDE filter), which is then followed by the low-thickness projection.}
	\label{fig:complete-workflow}
\end{figure}

Finally, we gather all the methods considered in this work and arrange them within the optimization loop, as shown in Fig. \ref{fig:complete-workflow}. This includes significant modifications within the boundary value problem (BVP) stage and the regularization stage. SIMP-based low-thickness suppression, described in Section \ref{sec:simp}, is part of the matrix assembly step, as in standard SIMP. Whereas the DGI projection plays a complementary role to density filtering, hence it directly follows it. Projection-based low-thickness suppression, described in Section \ref{sec:lt-proj}, is the final step in the regularization chain. For further clarity, the associated notation for the density variables is shown along the corresponding steps of the optimization loop.

As for the projection-based low-thickness suppression, we introduce a continuation scheme for the DGI projection sharpness $\hat \beta$. The continuation for $\hat \beta$ starts simultaneously with the continuation for the sharpness of the projection-based low-thickness suppression $\bar \beta$. Thus, the updated continuation workflow is outlined as follows:

\begin{enumerate}
	\item Start with $p = p_{\rm init} = 1$ and $\bar\beta = \bar\beta_{\rm init} = 1$
	\item Perform in every iteration:
	\begin{equation}
		p^{I+1} = c_p p^I
	\end{equation}
	and continue until $p = p_{\rm max} = 3$.
	\item Once $p = p_{\rm max} = 3$, perform in every iteration:
	\begin{equation}
		\hat\beta^{I+1} = \hat c_\beta \hat\beta^I
	\end{equation}
	\begin{equation}
		\bar\beta^{I+1} = \bar c_\beta \bar\beta^I
	\end{equation}
	and continue individually until $\hat\beta = \hat\beta_{\rm max}$ and $\bar\beta = \bar\beta_{\rm max}$, respectively.
\end{enumerate}

\subsection{Numerical example}

In the following numerical study, we employ the complete regularization procedure as shown in the previous section \ref{sec:workflow}. As the numerical study in Fig. \ref{fig:lt-study} has already shown the influence of varying low-thickness thresholds $\rho_{\rm low}$, there is no particular advantage in conducting such a study again, hence we focus only on the $\rho_{\rm low} = 0.1$ case. What is crucial, on the other hand, is to investigate the combination of various filter radii and the final sharpness parameters for DGI projection $\hat \beta_{\max}$, which are essentially complementary. Thus, variation of these two parameters is the main focus of this section. Besides this, a reasonable choice for the new continuation parameter $\hat c_\beta$ has to be made. To keep it straightforward, we choose the same continuation value for both DGI projection and low-thickness projection $\hat c_\beta = \bar c_\beta = 1.05$. Our additional numerical studies have shown that the method is only slightly sensitive to various combinations of these continuation parameters. Hence, such a study does not contribute significantly to this work, other than confirming the robustness of the complete approach. 

\begin{figure*}[bt]
	\centering
	\includegraphics[width=174mm]{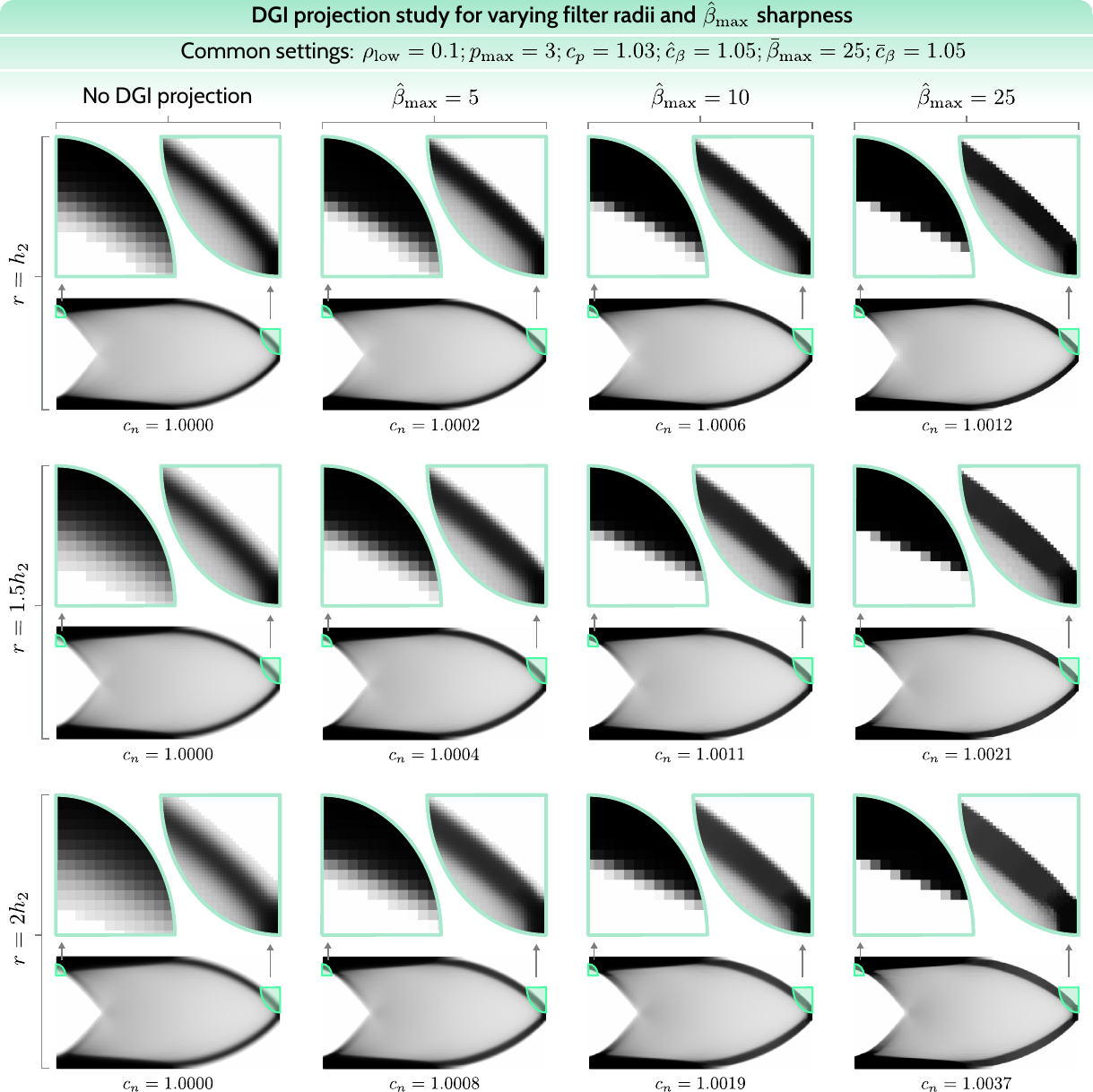}
	\caption{Final structures for the cantilever benchmark including DGI projection. The study focuses on the relation between the filter radius and the target projection sharpness $\hat \beta_{\max}$. Below each case, normalized objective values are shown.}
	\label{fig:dgi-study}
\end{figure*}

\begin{figure*}[bt]
	\centering
	\includegraphics[width=174mm]{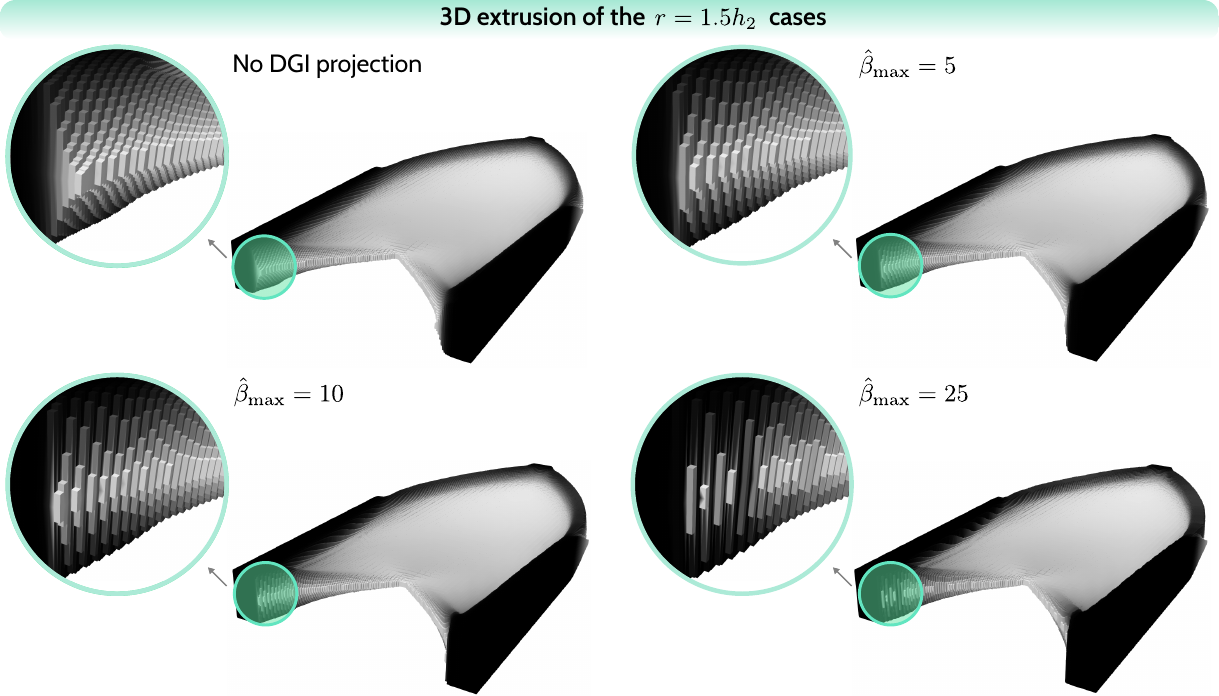}
	\caption{3D interpretation of the cases from the row $r = 1.5h_2$ from Fig. \ref{fig:dgi-study}, including close-up views of a representative region. The 3D interpretation is created by a simple extrusion of the finite element cells proportional to the $\bar \rho$ values. This simple 3D rendering is purely for visualization.}
	\label{fig:dgi-study-3d}
\end{figure*}

Thus, in Fig. \ref{fig:dgi-study}, we show the final cantilever structures for the $r/\hat \beta_{\max}$ study, including reference cases without DGI projection for each filter radius $r$. Since the main areas of interest are the structural edges, for each case two close-up views of the most representative regions are added. In addition, normalized compliance values are shown for performance considerations. For each row, i.e., each filter radius $r$, the normalization factor is the compliance of the reference case in that row. For instance, for the $r/\hat \beta_{\max} = 1.5h_2 / 10$ case, the normalized compliance is calculated by dividing its total compliance with the total compliance of the reference case (no DGI projection) with $r = 1.5h_2$: $c_n = 7.63657 \times 10^{-5} / 7.62813 \times 10^{-5} \approx 1.0011$. Such normalization facilitates a better comparison of the influence of DGI projection on performance. A complementary Fig. \ref{fig:dgi-study-3d} shows a 3D representation of selected structures, in analogy to Fig. \ref{fig:lt-edges}, including a close-up of a representative region. The selected structures for the 3D view correspond to the row $r = 1.5h_2$ in Fig. \ref{fig:dgi-study}.

Considering the views of the complete structures, no visible variations are noted within the interior part. Only by inspecting the close-up views is the influence of DGI projection visible at the structural edges. For all of the studied filter radii, the sharpness values of $\hat \beta_{\max} = 10$ and $\hat \beta_{\max} = 25$ render near-optimal edge sharpness, considering the discussion in Section \ref{sec:sharpness}, suggesting an ideal value to be within this range. Thus, the values of $\hat \beta_{\max} = 10$ and $\hat \beta_{\max} = 25$ are acceptable, while the value of $\hat \beta_{\max} = 5$ does not provide sufficient deblurring. Moreover, this observation also suggests that DGI projection is fairly independent of the filter radius, because as the filter radius increases, the neighborhood $N_r(e)$ in DGI projection expands correspondingly, partially mitigating the effect of the blurring filter. Naturally, the differences in edge sharpness are still visible up to $\hat \beta_{\max} = 10$, as, strictly speaking, the filter radius independence is not completely guaranteed. 

DGI projection barely affects flat regions in the structure's interior. However, fairly steep variations in density (higher density gradient), as seen in the second close-up view, are affected by the DGI projection. On the other hand, the same regions are also significantly smoothed by a blurring filter, justifying a locally stronger influence of DGI projection. Crucially, DGI projection has no significant influence on the objective value, with the largest relative increase in the objective value of $0.37\%$ for the most aggressive case $r/\hat \beta_{\max} = 2 h_2 / 25$. This suggests that DGI projection is a non-invasive regularization tool, which does not significantly affect the design space. 

\begin{figure}[tb]
	\centering
	\includegraphics[width=84mm]{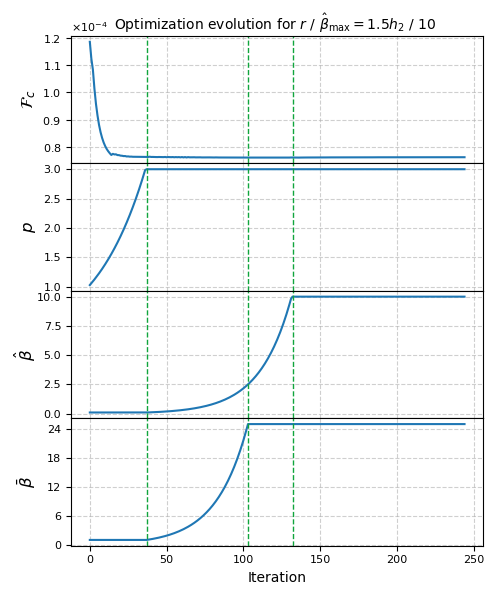}
	\caption{Evolution of the objective function and the continuation parameters - penalty $p$, DGI projection sharpness $\hat \beta$ and low-thickness projection sharpness $\bar \beta$ - for the case $r/\hat \beta_{\max} = 1.5h_2 / 10$. Vertical striped lines mark the transitions between the continuation phases. Note that $\hat \beta_{\rm init} = 0.1$ and $\bar \beta_{\rm init} = 1$.}
	\label{fig:evolution}
\end{figure}

In the following, we pick the case $r/\hat \beta_{\max} = 1.5 h_2 / 10$ and plot the evolution of the objective function together with the continuation parameters, see Fig. \ref{fig:evolution}. We can distinguish three main stages: the $p$-continuation, a combined $\hat\beta$- and $\bar\beta$-continuation (each of varying duration), and a fine-tuning stage, where all parameters remain constant, i.e., which starts after both $\hat\beta$ and $\bar\beta$ stop being updated. With our choice of the update parameters $c_p$, $\hat c_\beta$ and $\bar c_\beta$, the $p$-continuation takes around 40 iterations. During this stage, a rough design forms and the objective reaches a near-final value. During $\hat\beta$- and $\bar\beta$-continuation, the structural edges are clearly formed and any low-thickness densities are eliminated. The fine-tuning stage shows marginal changes, barely visible upon direct comparison.

\section{Conclusions and outlook} \label{sec:concl}

This work addressed two significant challenges in VTTO: the formation of undesirable low thickness regions and the blurring of structural edges caused by necessary regularization filters.

First, we presented a robust, combined approach to eliminate low thickness regions. This method merges a SIMP-based penalization for densities below a threshold $\rho_{\rm low}$ with an updated, stable projection-based suppression. Numerical studies demonstrated that this combined strategy, using a sequential continuation, effectively removes almost all low thickness densities, proving more robust than using either the penalization or projection method alone.

Second, the main contribution of this work is the new density-gradient-informed (DGI) projection, which is designed to solve the problem of blurred structural edges. This projection operates after the blurring filter. Its key feature is that it uses local information about the density gradient (specifically, the local density variation $d_r$) to adjust the projection sharpness $\hat\beta$. As a result, the DGI projection selectively targets and sharpens the structural edges, where the density gradient is high. Regions in the structure's interior, which have a low density gradient, are minimally affected.

The numerical examples showed that the DGI projection successfully deblurs the edges, restoring a natural and sharp, "anti-aliased" transition from solid to void. This avoids the artifact of edges being fixed at the low thickness threshold $\rho_{\rm low}$. A critical finding is that this significant improvement in edge definition is achieved with a negligible impact on the final compliance objective. The method is non-invasive and provides a valuable regularization tool.

For future work, the proposed complete regularization workflow should be applied to more complex problems where sharp structural edges are critical. A promising direction is to investigate this method in stress-constrained optimization or for problems with singularity features (like the L-beam), where blurred or low-thickness edges can cause inaccurate stress amplification.

\bmhead{Acknowledgements}
Funded by the European Union. Views and opinions expressed are however those of the author(s) only and do not necessarily reflect those of the European Union or the European Research Council Executive Agency. Neither the European Union nor the granting authority can be held responsible for them. This work is supported by the European Research Council (ERC) under the Horizon Europe program (Grant-No. 101052785, project: SoftFrac).
\begin{figure}[h!]
	\centering
	\includegraphics[width=0.3\textwidth]{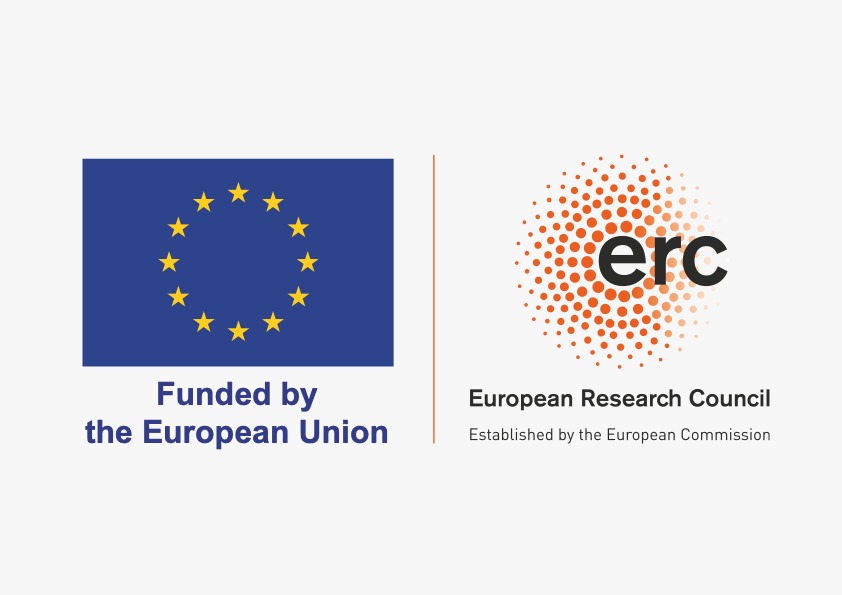}
\end{figure}
\section*{Declarations}
\bmhead{Author contributions}
Conceptualization: Gabriel Stankiewicz; Methodology: Gabriel Stankiewicz; Implementation: Gabriel Stankiewicz and Chaitanya Dev; Discussions and improvement of the methodology: all authors; Writing - original draft preparation: Gabriel Stankiewicz; Writing - review and editing: all authors; Funding acquisition: Paul Steinmann, Resources: Paul Steinmann, Supervision: Paul Steinmann.
\bmhead{Funding}
This work is supported by the European Research Council (ERC) under the Horizon Europe program (Grant-No. 101052785, project: SoftFrac).
\bmhead{Data availability}
The version of the code, the executable, the parameter settings and the result files are available from the corresponding author upon request.
\bmhead{Conflict of interest}
On behalf of all authors, the corresponding author states that there is no conflict of interest.
\bmhead{Ethical approval}
This study does not involve human participants, animal subjects, or ethical concerns requiring institutional approval.
\bmhead{Replication of results}
All the presented methodology is implemented in C++ utilizing the finite element library deal.II \citep{bangerth2007deal, arndt2021deal}. The version of the code, the executable, the parameter settings and the result files are available from the corresponding author upon request.
\bibliography{references}

@article{michell1904lviii,
  title={LVIII. The limits of economy of material in frame-structures},
  author={Michell, Anthony George Maldon},
  journal={The London, Edinburgh, and Dublin Philosophical Magazine and Journal of Science},
  volume={8},
  number={47},
  pages={589--597},
  year={1904},
  publisher={Taylor \& Francis}
}

@article{bendsoe1988generating,
  title={Generating optimal topologies in structural design using a homogenization method},
  author={Bends{\o}e, Martin Philip and Kikuchi, Noboru},
  journal={Computer methods in applied mechanics and engineering},
  volume={71},
  number={2},
  pages={197--224},
  year={1988},
  publisher={Elsevier}
}

@article{bendsoe1989optimal,
  title={Optimal shape design as a material distribution problem},
  author={Bends{\o}e, Martin P},
  journal={Structural optimization},
  volume={1},
  pages={193--202},
  year={1989},
  publisher={Springer}
}

@article{wang2003level,
  title={A level set method for structural topology optimization},
  author={Wang, Michael Yu and Wang, Xiaoming and Guo, Dongming},
  journal={Computer methods in applied mechanics and engineering},
  volume={192},
  number={1-2},
  pages={227--246},
  year={2003},
  publisher={Elsevier}
}

@article{rossow1973finite,
  title={A finite element method for the optimal design of variable thickness sheets},
  author={Rossow, MP and Taylor, JE},
  journal={Aiaa Journal},
  volume={11},
  number={11},
  pages={1566--1569},
  year={1973}
}

@article{sigmund2016non,
  title={On the (non-) optimality of Michell structures},
  author={Sigmund, Ole and Aage, Niels and Andreassen, Erik},
  journal={Structural and Multidisciplinary Optimization},
  volume={54},
  pages={361--373},
  year={2016},
  publisher={Springer}
}

@article{kandemir2018topology,
  title={Topology optimization of 2.5 D parts using the SIMP method with a variable thickness approach},
  author={Kandemir, Volkan and Dogan, Oguz and Yaman, Ulas},
  journal={Procedia Manufacturing},
  volume={17},
  pages={29--36},
  year={2018},
  publisher={Elsevier}
}

@article{yarlagadda2022solid,
  title={Solid isotropic material with thickness penalization--A 2.5 D method for structural topology optimization},
  author={Yarlagadda, Tejeswar and Zhang, Zixin and Jiang, Liming and Bhargava, Pradeep and Usmani, Asif},
  journal={Computers \& Structures},
  volume={270},
  pages={106857},
  year={2022},
  publisher={Elsevier}
}

@article{giele2021approaches,
  title={On approaches for avoiding low-stiffness regions in variable thickness sheet and homogenization-based topology optimization},
  author={Giele, Reinier and Groen, Jeroen and Aage, Niels and Andreasen, Casper Schousboe and Sigmund, Ole},
  journal={Structural and Multidisciplinary Optimization},
  volume={64},
  number={1},
  pages={39--52},
  year={2021},
  publisher={Springer}
}

@article{pozo2023minimum,
  title={Minimum-thickness method for 2.5 D topology optimization applied to structural design},
  author={Pozo, Sebastian and Golecki, Thomas and Gomez, Fernando and Carrion, Juan and Spencer Jr, Billie F},
  journal={Engineering Structures},
  volume={286},
  pages={116065},
  year={2023},
  publisher={Elsevier}
}

@inproceedings{endress2023designing,
  title={Designing variable thickness sheets for additive manufacturing using topology optimization with grey-scale densities},
  author={Endress, Felix and Zimmermann, Markus},
  booktitle={International Conference on Additive Manufacturing in Products and Applications},
  pages={63--76},
  year={2023},
  organization={Springer}
}

@article{groen2018homogenization,
  title={Homogenization-based topology optimization for high-resolution manufacturable microstructures},
  author={Groen, Jeroen P and Sigmund, Ole},
  journal={International Journal for Numerical Methods in Engineering},
  volume={113},
  number={8},
  pages={1148--1163},
  year={2018},
  publisher={Wiley Online Library}
}

@article{larsen2018optimal,
  title={Optimal truss and frame design from projected homogenization-based topology optimization},
  author={Larsen, SD and Sigmund, O and Groen, JP},
  journal={Structural and Multidisciplinary Optimization},
  volume={57},
  pages={1461--1474},
  year={2018},
  publisher={Springer}
}

@article{li2018topology,
  title={Topology optimization for concurrent design of structures with multi-patch microstructures by level sets},
  author={Li, Hao and Luo, Zhen and Gao, Liang and Qin, Qinghua},
  journal={Computer Methods in Applied Mechanics and Engineering},
  volume={331},
  pages={536--561},
  year={2018},
  publisher={Elsevier}
}

@article{banh2019topology,
  title={Topology optimization of multi-directional variable thickness thin plate with multiple materials},
  author={Banh, Thanh T and Lee, Dongkyu},
  journal={Structural and Multidisciplinary Optimization},
  volume={59},
  pages={1503--1520},
  year={2019},
  publisher={Springer}
}

@inproceedings{zhao2017stress,
  title={Stress-constrained thickness optimization for shell object fabrication},
  author={Zhao, Haiming and Xu, Weiwei and Zhou, Kun and Yang, Yin and Jin, Xiaogang and Wu, Hongzhi},
  booktitle={Computer Graphics Forum},
  volume={36},
  number={6},
  pages={368--380},
  year={2017},
  organization={Wiley Online Library}
}

@article{meng2022shape,
  title={Shape--thickness--topology coupled optimization of free-form shells},
  author={Meng, Xianchuan and Xiong, Yulin and Xie, Yi Min and Sun, Yuxin and Zhao, Zi-Long},
  journal={Automation in Construction},
  volume={142},
  pages={104476},
  year={2022},
  publisher={Elsevier}
}

@article{nguyen2022multi,
  title={Multi-material gradient-free proportional topology optimization analysis for plates with variable thickness},
  author={Nguyen, Minh Ngoc and Bui, Tinh Quoc},
  journal={Structural and Multidisciplinary Optimization},
  volume={65},
  number={3},
  pages={75},
  year={2022},
  publisher={Springer}
}

@article{sorensen2014dmto,
  title={DMTO--a method for discrete material and thickness optimization of laminated composite structures},
  author={S{\o}rensen, S{\o}ren N and S{\o}rensen, Ren{\'e} and Lund, Erik},
  journal={Structural and Multidisciplinary Optimization},
  volume={50},
  number={1},
  pages={25--47},
  year={2014},
  publisher={Springer}
}

@article{stegmann2005discrete,
  title={Discrete material optimization of general composite shell structures},
  author={Stegmann, Jan and Lund, Erik},
  journal={International Journal for Numerical Methods in Engineering},
  volume={62},
  number={14},
  pages={2009--2027},
  year={2005},
  publisher={Wiley Online Library}
}

@article{kashanian2021novel,
  title={A novel method for concurrent thickness and material optimization of non-laminate structures},
  author={Kashanian, Kiarash and Kim, Il Yong},
  journal={Structural and Multidisciplinary Optimization},
  volume={64},
  number={3},
  pages={1421--1437},
  year={2021},
  publisher={Springer}
}

@article{sjolund2018new,
  title={A new thickness parameterization for Discrete Material and Thickness Optimization},
  author={Sj{\o}lund, JH and Peeters, D and Lund, E},
  journal={Structural and Multidisciplinary Optimization},
  volume={58},
  pages={1885--1897},
  year={2018},
  publisher={Springer}
}

@article{kim2021generalized,
  title={Generalized optimality criteria method for topology optimization},
  author={Kim, Nam H and Dong, Ting and Weinberg, David and Dalidd, Jonas},
  journal={Applied Sciences},
  volume={11},
  number={7},
  pages={3175},
  year={2021},
  publisher={MDPI}
}

@article{lazarov2011filters,
  title={Filters in topology optimization based on Helmholtz-type differential equations},
  author={Lazarov, Boyan Stefanov and Sigmund, Ole},
  journal={International journal for numerical methods in engineering},
  volume={86},
  number={6},
  pages={765--781},
  year={2011},
  publisher={Wiley Online Library}
}

@article{wang2011projection,
  title={On projection methods, convergence and robust formulations in topology optimization},
  author={Wang, Fengwen and Lazarov, Boyan Stefanov and Sigmund, Ole},
  journal={Structural and multidisciplinary optimization},
  volume={43},
  number={6},
  pages={767--784},
  year={2011},
  publisher={Springer}
}

@article{bangerth2007deal,
  title={deal. II—a general-purpose object-oriented finite element library},
  author={Bangerth, Wolfgang and Hartmann, Ralf and Kanschat, Guido},
  journal={ACM Transactions on Mathematical Software (TOMS)},
  volume={33},
  number={4},
  pages={24--es},
  year={2007},
  publisher={ACM New York, NY, USA}
}

@article{arndt2021deal,
  title={The deal. II finite element library: Design, features, and insights},
  author={Arndt, Daniel and Bangerth, Wolfgang and Davydov, Denis and Heister, Timo and Heltai, Luca and Kronbichler, Martin and Maier, Matthias and Pelteret, Jean-Paul and Turcksin, Bruno and Wells, David},
  journal={Computers \& Mathematics with Applications},
  volume={81},
  pages={407--422},
  year={2021},
  publisher={Elsevier}
}

@article{geoffroy2022coupled,
	title={Coupled optimization of macroscopic structures and lattice infill},
	author={Geoffroy-Donders, Perle and Allaire, Gr{\'e}goire and Michailidis, Georgios and Pantz, Olivier},
	journal={International Journal for Numerical Methods in Engineering},
	volume={123},
	number={13},
	pages={2963--2985},
	year={2022},
	publisher={Wiley Online Library}
}

@article{stankiewicz2025novel,
	title={A novel multi-thickness topology optimization method for balancing structural performance and manufacturability},
	author={Stankiewicz, Gabriel and Dev, Chaitanya and Steinmann, Paul},
	journal={arXiv preprint arXiv:2507.19388},
	year={2025}
}

@article{stankiewicz2025configurational,
	title={Configurational-force-driven adaptive refinement and coarsening in topology optimization},
	author={Stankiewicz, Gabriel and Dev, Chaitanya and Steinmann, Paul},
	journal={Structural and Multidisciplinary Optimization},
	volume={68},
	number={8},
	pages={1--15},
	year={2025},
	publisher={Springer}
}

@article{stankiewicz2022geometrically,
	title={Geometrically nonlinear design of compliant mechanisms: Topology and shape optimization with stress and curvature constraints},
	author={Stankiewicz, Gabriel and Dev, Chaitanya and Steinmann, Paul},
	journal={Computer Methods in Applied Mechanics and Engineering},
	volume={397},
	pages={115161},
	year={2022},
	publisher={Elsevier}
}

\end{document}